 \documentclass[smallabstract,smallcaptions]{dccpaper}

\usepackage{epsfig}
\usepackage{amsmath}
\usepackage{amssymb}
\usepackage{bbm}
\usepackage{color}
\usepackage{url}
\usepackage{cite}
\usepackage{subfig}
\usepackage{epstopdf}
\DeclareRobustCommand{\stirling}{\genfrac\{\}{0pt}{}}

\newlength{\figurewidth}
\newlength{\smallfigurewidth}

\begin{document}

\title
{\large
\textbf{The Twelvefold Way of Non-Sequential Lossless Compression}
}

\author{%
Illinois Information Theory Students,$^\dagger$ \thanks{$^\dagger$ Illinois Information Theory Students (FA20): Taha Ameen ur Rahman, Alton S. Barbehenn, Xinan Chen, Hassan Dbouk, James A. Douglas, Yuncong Geng, Ian George, John B. Harvill, Sung Woo Jeon, 	Kartik K. Kansal, Kiwook Lee, Kelly A. Levick, Bochao Li, Ziyue Li, Yashaswini Murthy, Adarsh Muthuveeru-Subramaniam, S. Yagiz Olmez, Matthew J. Tomei, Tanya Veeravalli, Xuechao Wang, Eric A. Wayman, Fan Wu, Peng Xu, Shen Yan, Heling Zhang, Yibo Zhang, Yifan Zhang, and Yibo Zhao.} Sourya Basu, and Lav R. Varshney\\[0.5em]
{\small\begin{minipage}{\linewidth}\begin{center}
\begin{tabular}{ccc}
University of Illinois at Urbana-Champaign\\
\end{tabular}
\end{center}\end{minipage}}
}

\maketitle
\thispagestyle{empty}

\begin{abstract}
Many information sources are not just sequences of distinguishable symbols but rather have invariances governed by alternative counting paradigms such as permutations, combinations, and partitions.  We consider an entire classification of these invariances called the twelvefold way in enumerative combinatorics and develop a method to characterize lossless compression limits.  Explicit computations for all twelve settings are carried out for i.i.d.\ uniform and Bernoulli distributions.  Comparisons among settings provide quantitative insight.
\end{abstract}

\section{Introduction}\label{sec: introduction}

Several frontiers of data science are producing huge amounts of data that are not simple sequences of distinguishable symbols, but have certain equivalence classes for patterns of symbols within which lossless representation does not require unique indices.  As a simple example, the ordering of sequences of scientific data records are often irrelevant and so only the histogram or type class must be represented \cite{VarshneyG2006,VarshneyG2006b,Reznik2011,GriponRSG2012,Steinruecken2015,OrlitskySVZ2006}.  Similar representation problems arise for biological data \cite{GinartHZNCST2018}, social graph structure \cite{ChoiS2012}, neural network architecture \cite{BasuV2017}, and elsewhere due to the functional process generating or using the data.  In fact, these \emph{non-sequential} information sources are prevalent in the science of information \cite{Szpankowski2012} and may be studied implicitly via group theory \cite{Lim2012}. Some non-sequential sources also arise in computer science via trace theory, which can be studied explicitly using interchange entropy \cite{Savari2004}.

Here we draw on enumerative combinatorics to explicitly find entropy bounds for a large class of possible kinds of non-sequential sources defined using functions between two finite sets.  We specifically consider the twelvefold way in combinatorics due to Gian-Carlo Rota as a general classification of non-sequentiality that encompasses counting of permutations, combinations, multisets, and partitions (see \cite{Stanley1997} for attribution and importance in applications), and therefore significantly generalizes from just  irrelevance of order. The key mathematical step in information-theoretic analysis beyond known combinatorics is to characterize how probability distributions collapse under various invariances. 

Sec.~\ref{sec: problem_settings}  reviews the twelvefold way in combinatorics and introduces the high-level entropy computation problem.  Later sections carry out this program for the twelve possible settings. Sec.~\ref{sec:conclusion} concludes with avenues for future work.

\section{Problem Settings}\label{sec: problem_settings}

The twelvefold way in combinatorics is a collection of twelve related problems that involve enumeration of different equivalence classes of functions between two finite sets with further restrictions on the functions~\cite{Stanley1997}. There are three possible restrictions for the functions: i) no restrictions, ii) injective, or iii) surjective. Further, for the set of functions from $\mathcal{N}=\{1,\ldots,n\}$ to $\mathcal{X}=\{1,\ldots,x\}$, there are four possible equivalence relations: i) equality, ii) equality up to a permutation of $\mathcal{N}$, iii) equality up to a permutation of $\mathcal{X}$, iv) equality up to permutations of $\mathcal{N}$ and $\mathcal{X}$. Considering the three possible restrictions on the functions and four possible equivalence relations on the set of functions, we get twelve different counting problems.

The significance of these twelve settings in the context of lossless compression can be understood from one popular interpretation of function $f:\mathcal{N}\mapsto \mathcal{X}$ as filling $|\mathcal{N}|=n$ bins using $|\mathcal{X}|=x$ colors. Sequential data has been the most widely studied datatype for lossless compression and can be interpreted as a realization of $f$ as $(f(1),\ldots,f(n))$ for no restrictions on $f$ with equality of function realizations as equivalence relation. Much is known about the entropy of such sequential data. To this end, we see these twelve settings as a generalization of sequential labeled data. 

Now we develop an algorithm to compute entropy in these settings, so as to quantify the differences among them; later we describe the underlying stochastic process considered.
For given $\mathcal{N}$, $\mathcal{X}$, and restrictions on $f$, we compute the entropy of $(f(1),\ldots,f(n))$, denoted $H(f(1),\ldots,f(n))$, as follows:
\begin{enumerate}
    \item Generate $X^n = (X_1,\ldots,X_n)$ i.i.d.\ from probability mass function $P$ over $\mathcal{X}$.
    \item Remove all the sequences that do not follow the restrictions on $f$.
    \item Renormalize the probability distribution of the sequences by dividing the probability of each sequence by the sum of the probabilities of the remaining sequences.
    \item Group the sequences based on the equivalence relations.
    \item Compute the entropy of the grouped probabilities.
\end{enumerate}

Thus, for a fixed $P$ over $\mathcal{X}$, the probability of a sequence ${X^n = (X_1,\ldots,X_n)}$ before grouping based on equivalence relations is proportional to
\begin{align}
    \left( \prod_{i}^n P(X_i) \right) \mathbbm{1}_{\{X^n \text{ is valid}\}},
\end{align}
where $\mathbbm{1}_{\{\}}$ is an indicator function and a sequence $X^n$ is valid if it follows the restrictions imposed on the function $f$. Then, these valid sequences are grouped according to the equivalence relations on $\mathcal{X}$ and $\mathcal{N}$. For concreteness we focus on two specific distributions that are perhaps the most basic: uniform distribution over $\mathcal{X}$ denoted $\mathcal{U}(\mathcal{X})$ and Bern($p$) distribution over $\mathcal{X}=\{0,1\}$. 
Now we detail the entropy computation in each of the cases.

\section{No condition on functions}\label{sec: no_condition_on_functions}
Let there be no constraint on the function $f$. We consider all three possible equivalence relations on the sets $\mathcal{N}$ and $\mathcal{X}$.
\paragraph{Equality of functions}\label{subsec: equality_1}
Here we consider the equivalence classes of equality of functions. This reduces to a sequence $X^n$ generated from an i.i.d.\ distribution $P$. Let us recall the entropy for $P \sim \mathcal{U}(\mathcal{X})$  and $\sim$Bern($p$).
For $\mathcal{U}(\mathcal{X})$ we have $H(f(1),\ldots,f(n)) = n\log_2{x}$ and for Bern($p$) distribution we have $H(f(1),\ldots,f(n)) = n h_2(p)$, where $h_2(p) = -p\log_2{p} - (1-p)\log_2{(1-p)}$ is the binary entropy.
\paragraph{Equality of functions up to a permutation of $\mathcal{N}$}
\label{sec: equality_up_to_N_1}
Here we consider the equivalence class of functions up to permutation of $\mathcal{N}$. First we consider $\mathcal{U}(\mathcal{X})$. There is no restriction on $f$ here, but we must group sequences based on their equivalence up to $\mathcal{N}$. Hence, any two sequences with the same composition of elements from $\mathcal{X}$ must be grouped together. Thus, any equivalence class of functions can be represented by its composition. Further, for any composition $(n_0,n_1,\ldots,n_{(x-1)})$, i.e.\ $n_i \geq 0$ occurrences of $i \in \mathcal{X}$, there are $n \choose {n_0n_1\cdots n_{(x-1)}}$ sequences. Thus, the entropy for this case is
\begin{equation}\label{eqn: any_n_u}
    H(f(1),\ldots,f(n)) = \sum_{(n_0,n_1,\ldots,n_{(x-1)}): \sum_{i}n_i = n} p(n_0,n_1,\ldots,n_{(x-1)})\log_2{\tfrac{1}{p(n_0,n_1,\ldots,n_{(x-1)})}},
\end{equation} 
where $p(n_0,n_1,\ldots,n_{(x-1)}) = \frac{1}{x^n} {n \choose {n_0..n_{(x-1)}}}$. This quantity may be expensive to compute, hence when required, we approximate $H(f(1),\ldots,f(n))$ with an upper bound: 
\begin{align}\label{eqn: any_n_u_ub}
    H(f(1),\ldots,f(n)) \leq \log_2{{x+n-1 \choose n}},
\end{align}
since there are at most ${x+n-1 \choose n}$ different equivalence classes in this case.

For the Bern($p$) case, the composition $(n_0,n_1)$ can be described using the number of ones in the sequence. There are ${n \choose i}$ sequences of length $n$ with $i>0$ ones, each occuring with probability $p^i(1-p)^{(n-i)}$. Thus, the entropy in this case is just:
\begin{equation}\label{eqn: any_n_b}
    H(f(1),\ldots,f(n)) = \sum_{i=0}^n p(i,n-i)\log_2{\tfrac{1}{p(i,n-i)}},
\end{equation} 
where $p(i,n-i) = {n \choose i} p^i(1-p)^{(n-i)}$.

\paragraph{Equality of functions up to a permutation of $\mathcal{X}$}\label{sec: equality_up_to_X_1}
Here we consider equivalence relations up to a permutation of $\mathcal{X}$. First let us take $P\sim\mathcal{U}(\mathcal{X})$. In this case, we start with grouping equivalent sequences. Suppose a sequence has $k$ different colors, then there are ${x^{\underline{k}} = x(x-1)\cdots (x-k+1)}$ possible ways to permute the colors. Hence, equivalence class of a sequence with $k$ colors have $x^{\underline{k}}$ elements in them. The probability of this equivalence class is given by $\frac{x^{\underline{k}}}{x^n}$. Note that not all sequences with $k$ colors belong to the same equivalence class. In fact, observe that there are $\stirling{n}{k}$ many equivalence classes with $k$ colored sequences, where $\stirling{n}{k} = \frac{1}{k!}\sum_{i=0}^k (-1)^i {k \choose i} (k-i)^n$ is Stirling's number of the second kind. Thus, the entropy can be computed as
\begin{equation}\label{eqn: any_x_u}
    H(f(1),\ldots,f(n)) = -\sum_{k=0}^x \stirling{n}{k} \frac{x^{\underline{k}}}{x^n} \log{\frac{x^{\underline{k}}}{x^n}}.
\end{equation}
Hence, the total number of equivalence classes is $\sum_{k=0}^x\stirling{n}{k}$; for ease of computation, the entropy $H(f(1),\ldots,f(n))$ can be upper-bounded as
\begin{align}\label{eqn: any_x_u_ub}
    H(f(1),\ldots,f(n)) &\leq \log_2{\sum_{k=0}^x\stirling{n}{k}}.
\end{align}

Now, let us consider the Bern($p$) case and have two different sub-cases of when $n$ is even or odd. When $n$ is odd, the equivalence class of each sequence consists of exactly two sequences, the sequence itself and the sequence obtained by flipping zeros and ones. Further, there are ${n \choose k}$ sequences with $k$ ones, hence the entropy is:
\begin{align}\label{eqn: any_x_b_odd}
    H(f(1),\ldots,f(n)) &= - \sum_{i=0}^{\frac{n-1}{2}} {n \choose k} p(k,n-k)\log_2{p(k,n-k)},
\end{align}
where $p(k,n-k) = \left( p^k(1-p)^{n-k} + (1-p)^k p^{n-k} \right)$.
When $n$ is even, we need to take care of additional base cases which makes the expression somewhat complicated. We omit this case for brevity since the main goal here is to understand the behavior of entropy for different cases.

\paragraph{Equality of functions up to permutations of $\mathcal{N}$ and $\mathcal{X}$}\label{sec: equality_up_to_N_and_X_1}
Here equivalence is with respect to permutation in $\mathcal{N}$ as well as $\mathcal{X}$. Computing the entropy for this case is somewhat involved, hence we will use an upper bound as an approximation. For the uniform distribution, we approximate the entropy using the upper bound
\begin{align}\label{eqn: any_nx_u_ub}
    H(f(1),\ldots,f(n)) &\leq \log_2{p_x(n+x)},
\end{align}
where $p_x(n)$ is the integer partition function \cite{HardyW1979} and $p_x(n+x)$ is the number of ways to divide the integer $n$ into less than or equal to $x$ parts.

For the Bern($p$) case, entropy computation is somewhat simpler and gives a better intuition. We take $n$ to be odd for simplicity. Then there are $\frac{n-1}{2}$ equivalence classes, where the $k$th equivalence class for $k \in [\frac{n-1}{2}]$ has either $k$ or $(n-k)$ ones in the sequences in each class. Hence the entropy can be computed as
\begin{align}\label{eqn: any_nx_b_odd}
    H(f(1),\ldots,f(n)) &= - \sum_{i=0}^{\frac{n-1}{2}} p(k,n-k) \log_2{p(k,n-k)},
\end{align}
where $p(k,n-k) = {n \choose k} \left( p^k(1-p)^{n-k} + (1-p)^k p^{n-k} \right)$.

\begin{figure}
    \centering
    \subfloat[]{{\includegraphics[width=6cm]{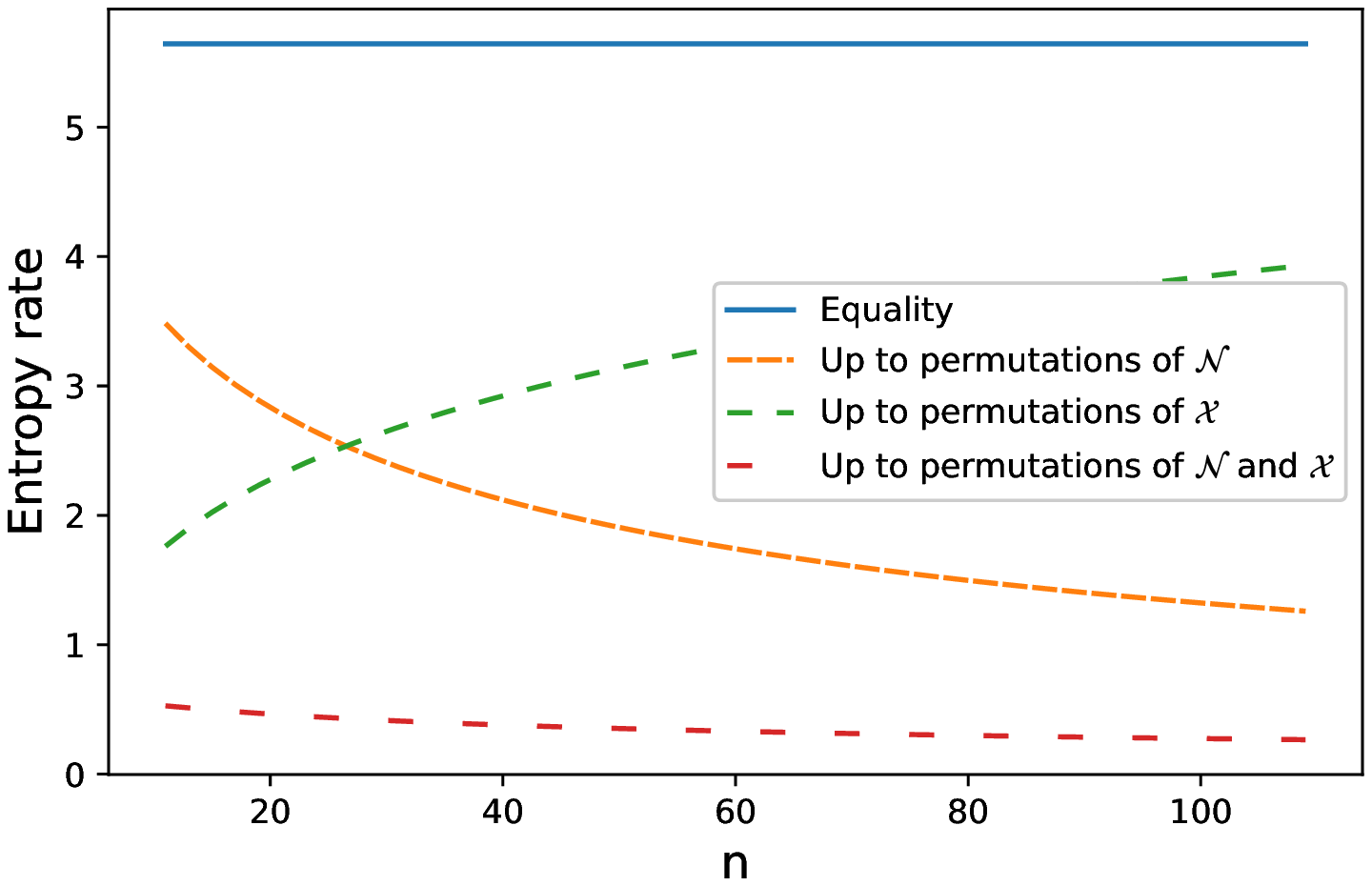} }\label{fig: N_U_C_1}}%
    \qquad
    \subfloat[]{{\includegraphics[width=6cm]{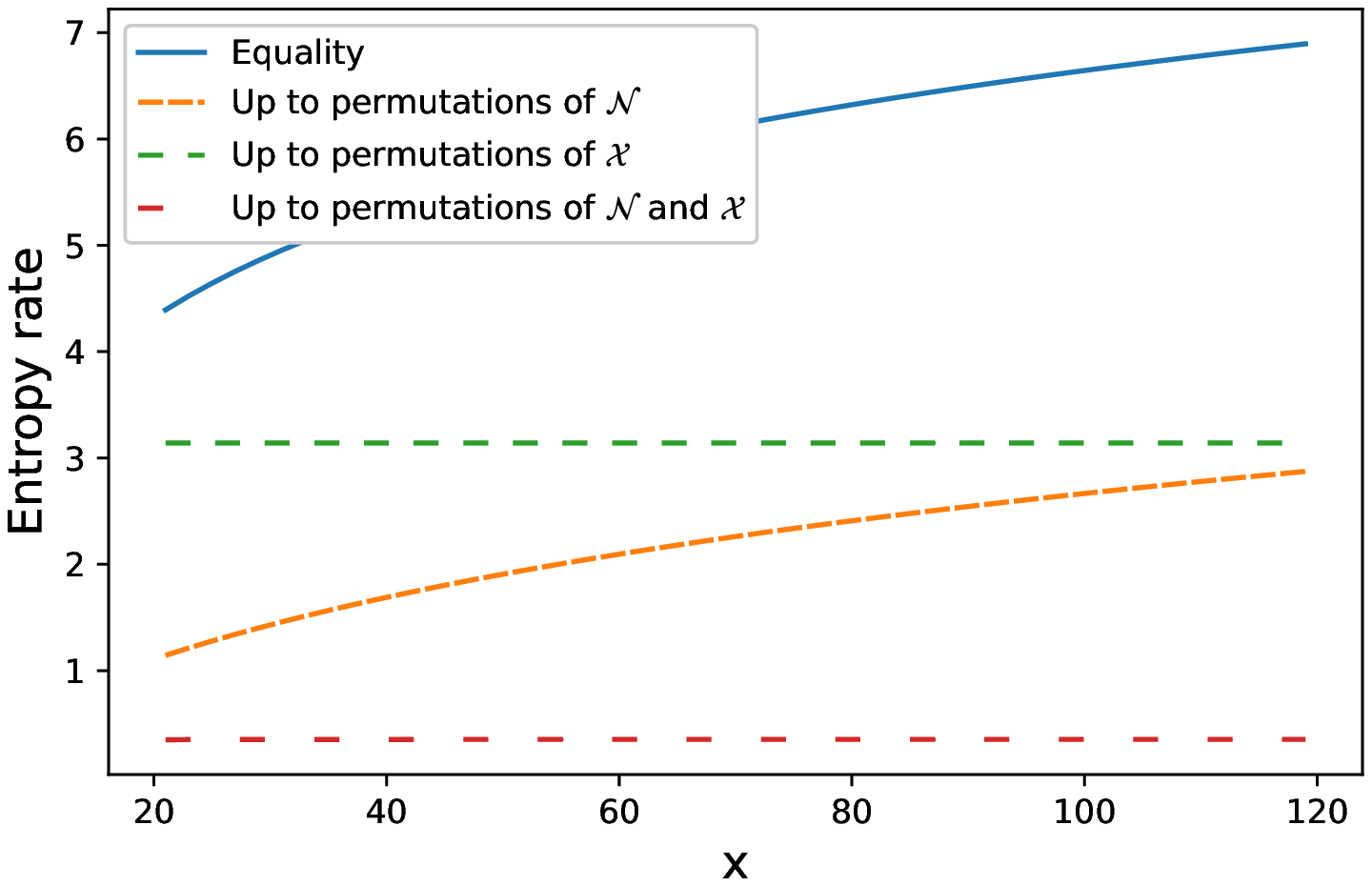} } \label{fig: X_U_C_1}}
    \qquad
    \subfloat[]{{\includegraphics[width=6cm]{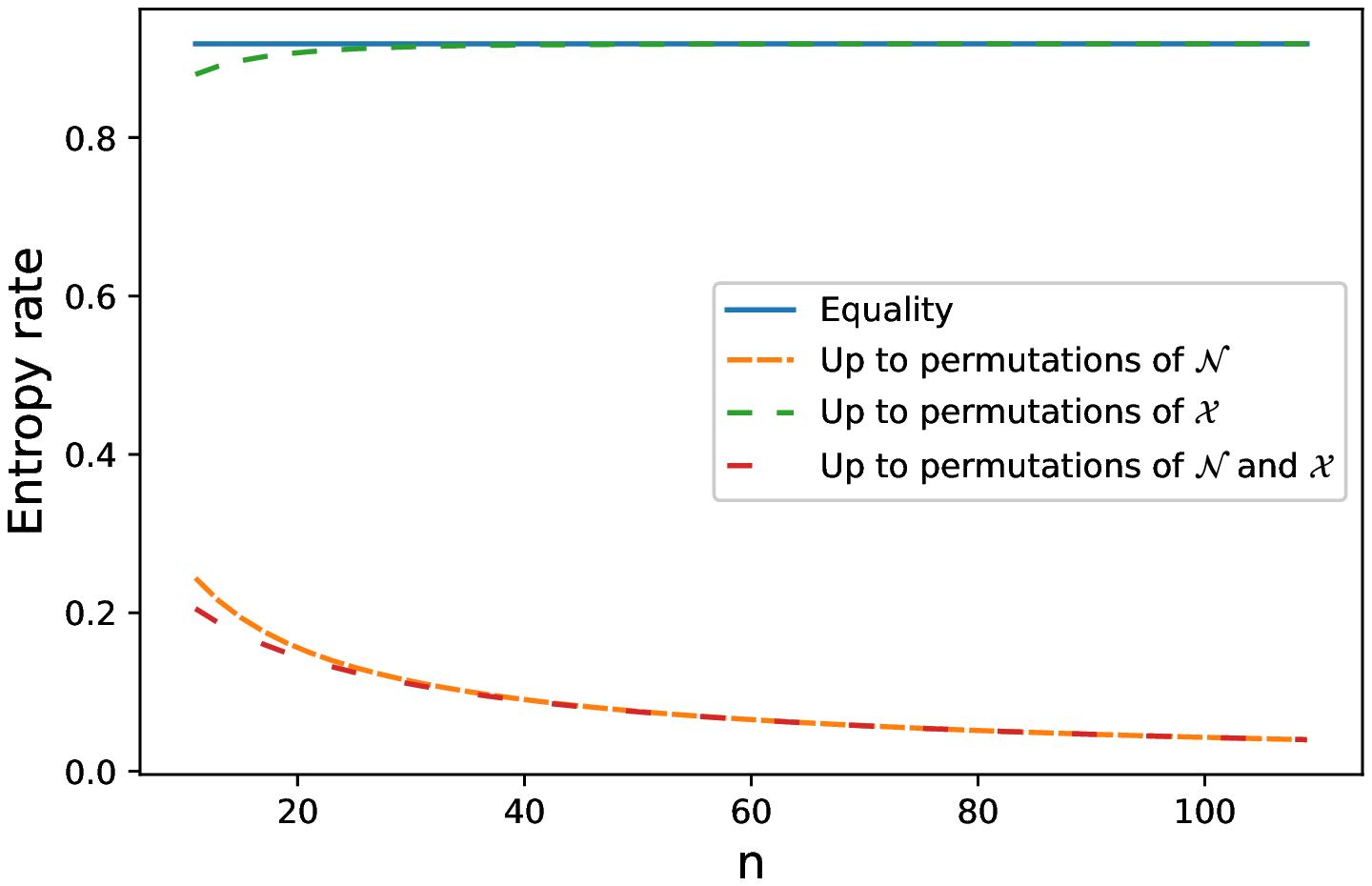} } \label{fig: N_B_C_1}}%
    \qquad
    \subfloat[]{{\includegraphics[width=6cm]{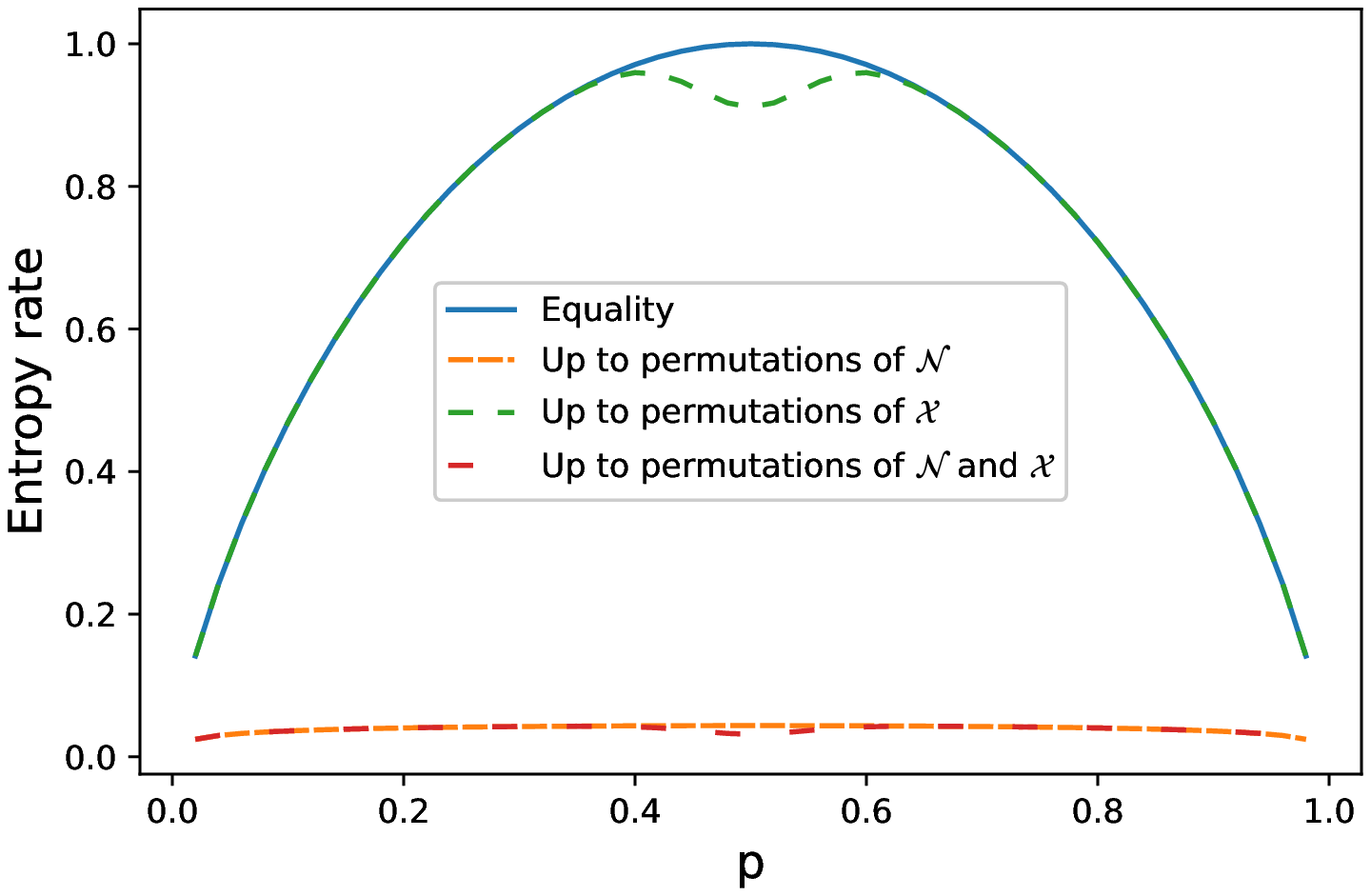} } \label{fig: P_B_C_1}}%
    \vspace{-3mm}
    \caption{Entropy rates for no condition on functions vs. parameters $n,x,$ and $p$ for uniform and Bernoulli distributions and various equivalence relations.}%
    \label{fig: comparison_1}
\end{figure}

\paragraph{Comparison of entropies for no condition on functions}\label{sec: comparison_1}
Entropy rate is simply defined as entropy per unit $n$. In Fig.~\ref{fig: N_U_C_1},~\ref{fig: X_U_C_1}, we plot entropy rates corresponding to $\log_2{x}$, \eqref{eqn: any_n_u_ub}, \eqref{eqn: any_x_u_ub}, \eqref{eqn: any_nx_u_ub} against $n$ with $x=50$ and $x$ with $n=50$ respectively. Importantly, we find \eqref{eqn: any_n_u_ub} and \eqref{eqn: any_x_u_ub} remain significantly below $\log_2{x}$. More details on the behavior of these expressions on $n$ and $x$ can also be noted from these plots. In Fig.~\ref{fig: N_B_C_1},~\ref{fig: P_B_C_1}, we plot the Bern($p$) counterparts of the above against $n$ with $p=1/3$ and $p$ with $n=100$ respectively to find a significant reduction in entropy rates for the cases with $\mathcal{N}$ equivalences.

\section{Injective functions}\label{sec: injective_functions}
Now, we will look at functions that are constrained to be injective. Here the implicit assumption is $n \leq x$ to have non-zero valid sequences.

\paragraph{Equality of functions}\label{subsec: equality_2}
Here we consider equivalence classes of equality of functions that are restricted to be injective.

First we take $\mathcal{U}(\mathcal{X})$. Note that in the case of uniform distribution, removing the invalid sequences and renormalizing the probability results in a uniform distribution over all the valid sequences and the number of valid sequences of length $n$ satisfying the injective conditions on $f$ can be computed to be $\frac{x!}{(x-n)!}$. Thus, the entropy is given by that of uniform distribution with alphabet of size $\frac{x!}{(x-n)!}$, i.e.\
\begin{align}
\label{eqn: inj_eq_u}
    H(f(1),\ldots,f(n)) = \log\frac{x!}{(x-n)!} = \sum_{k=1}^n\log({x-k+1}).
\end{align} 

For the Bern($p$) case, since we are considering injective functions, $n\leq x=2$, which makes this case somewhat redundant in terms of entropy computation. For $n=1$, the entropy is simply $H(f(1)) = h_2(p)$. For, $n=2$, the only two valid sequences are $(0,1)$ and $(1,0)$ which are equiprobable, hence $H(f(1),f(2))=1$.

\paragraph{Equality of functions up to a permutation of $\mathcal{N}$}\label{sec: equality_up_to_N_2}
First consider the uniform case. Because of the injective nature of functions, all positions in $\mathcal{N}$ are filled by different elements from $\mathcal{X}$, so there are $n!$ sequences in each equivalence class. Hence, all the equivalence classes are equiprobable. Further, there are $x \choose n$ equivalence classes, hence the entropy in this case is
\begin{align}\label{eqn: inj_n_u}
    H(f(1),\ldots,f(n)) &= \log{x \choose n}.
\end{align} 

The Bern($p$) case is somewhat trivial. Due to the constraint of injectivity, we have $n \leq x=2$. For $n=1$, the entropy is simply  $h_2(p)$. Whereas for $n=2$, we have only one possible sequence $(0,1)$ up to permutation of $\mathcal{N}$, hence the entropy is simply $0$.

\paragraph{Equality of functions up to a permutation of $\mathcal{X}$}\label{sec: equality_up_to_X_2}
This case as we will show has zero entropy. Note that we are working with injective functions, hence we have $n\leq x$. We also have equality of functions up to permutations of $\mathcal{X}$. Observe that all possible sequences lie in the same equivalence class, i.e.\ any sequence can be obtained by any other sequence simply by permuting $\mathcal{X}$. Thus, we have 
\begin{align}\label{eqn: inj_x_u}
    H(f(1),\ldots,f(n)) &= 0,
\end{align} 
for both the uniform and Bernoulli cases.

\begin{figure}
    \centering
    \subfloat[]{{\includegraphics[width=6cm]{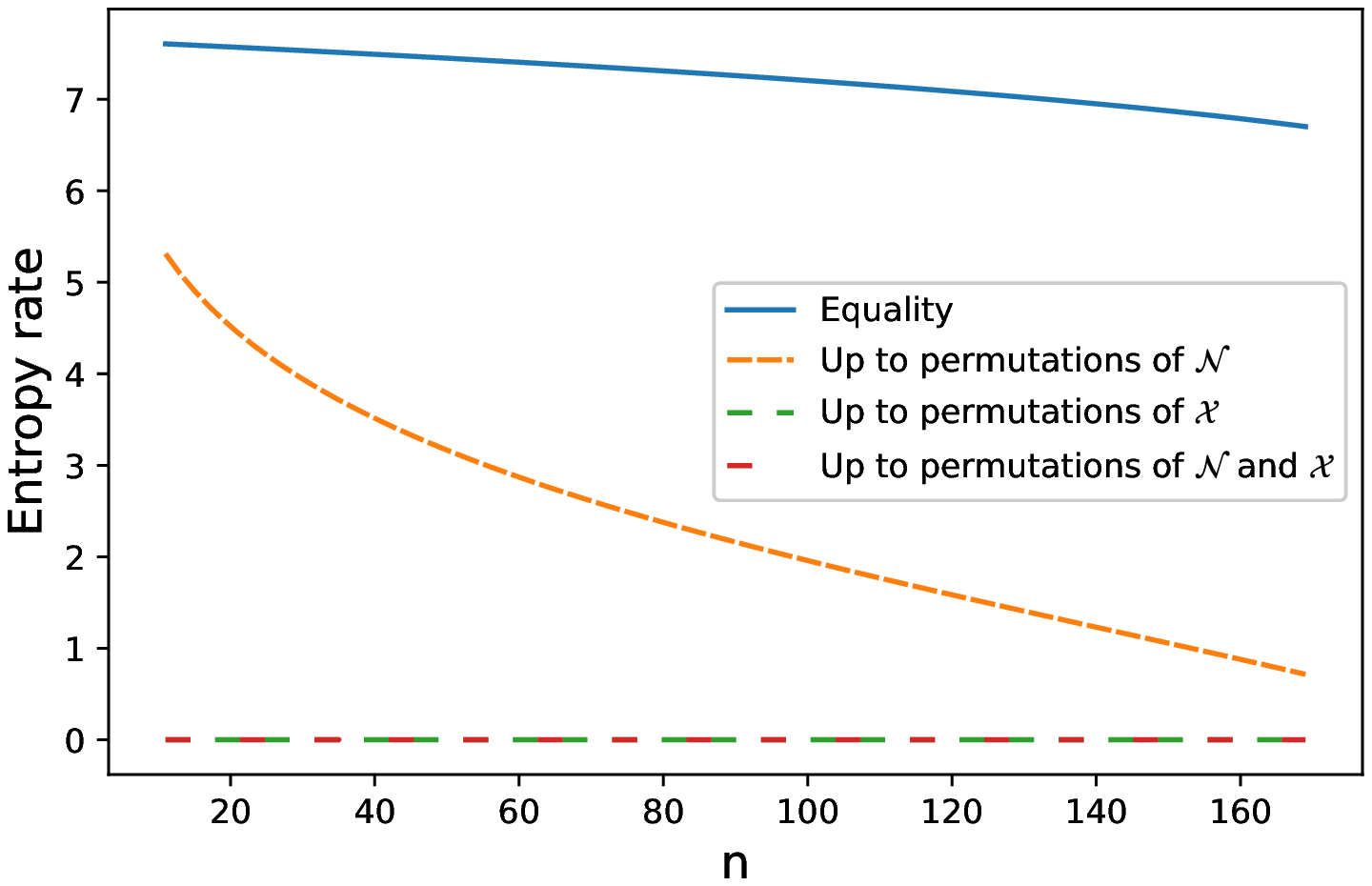} }\label{fig: N_U_C_2}}%
    \qquad
    \subfloat[]{{\includegraphics[width=6cm]{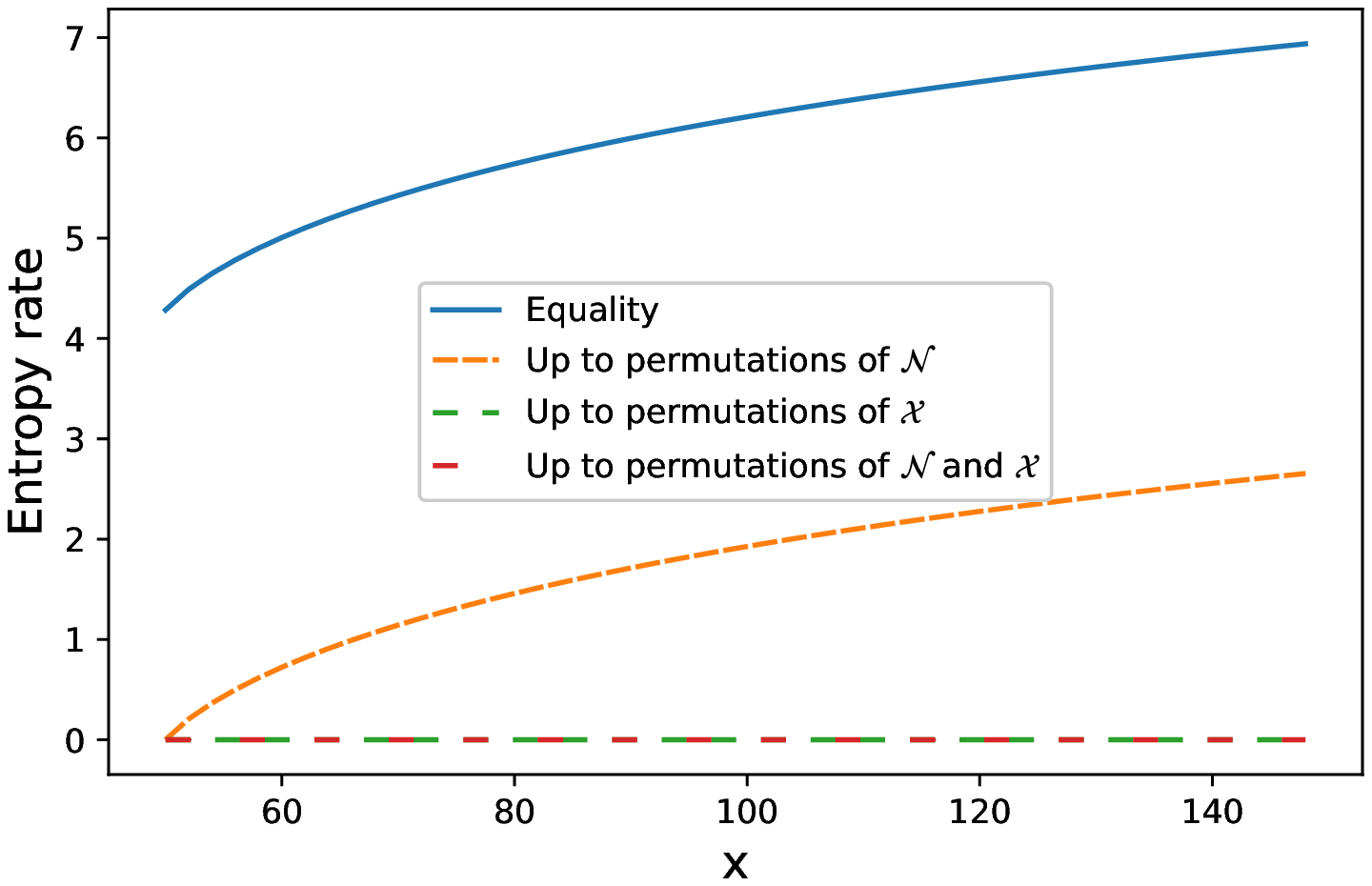} } \label{fig: X_U_C_2}}
    \vspace{-3mm}
    \caption{Entropy rates for injective functions vs. $n$ and $x$ for uniform distribution for various equivalence relations.}%
    \label{fig: comparison_2}
\end{figure}

\paragraph{Equality of functions up to permutations of $\mathcal{N}$ and $\mathcal{X}$}\label{sec: equality_up_to_N_and_X_2}
Here we have injective functions with equality up to permutations in both $\mathcal{X}$ and $\mathcal{N}$. 
Like the previous case, we have exactly one equivalence class which yields exactly zero entropy, i.e., 
\begin{align}\label{eqn: inj_nx_u}
    H(f(1),\ldots,f(n)) &= 0,
\end{align} 
for both uniform and Bern($p$) distributions.
\paragraph{Comparison of entropies for injective functions}\label{sec: comparison_2}
In Fig.~\ref{fig: N_U_C_2},~\ref{fig: X_U_C_2}, we plot entropy rates corresponding to \eqref{eqn: inj_eq_u}, \eqref{eqn: inj_n_u}, \eqref{eqn: inj_x_u}, \eqref{eqn: inj_nx_u} against $n$ with $x=200$ and $x$ with $n=50$ respectively. Importantly, we find that \eqref{eqn: inj_n_u} decreases with $n$ and increases with $x$ and \eqref{eqn: inj_x_u}, \eqref{eqn: inj_nx_u} remain at 0. 
The Bern($p$) cases for injective functions are trivial and hence not plotted here.
\section{Surjective functions}\label{sec: surjective_functions}
Here, we consider functions that are restricted to be surjective. Hence, we have the implicit assumption $x \leq n$ to have non-zero valid sequences.
\paragraph{Equality of functions}\label{subsec: equality_3}
With equivalence classes based on equality of surjective functions, we first consider $\mathcal{U}(\mathcal{X})$. We must compute the total number of valid sequences of length $n$ since we know that after renormalization of probabilities, they are all equiprobable. The total number of valid sequences can be computed to be $x! \stirling{n}{x}$. Thus, the entropy in this case is
\begin{align}\label{eqn: surj_eq_u}
    H(f(1),\ldots,f(n)) &= \log_2{\left( x! \stirling{n}{x}\right)},
\end{align} 
where $\stirling{n}{x}$ is the Stirling number of the second kind.

Now we consider when $P$ is Bern($p$). Since the functions are surjective, we have two invalid sequences: the all-zeros sequence and the all-ones sequence. Renormalizing the probabilities, we have the probability of a sequence with $k$ ones and $(n-k)$ zeros for $k \in \{1,\ldots,n-1\}$ as 
\begin{align}
    p_k = \frac{(1-p)^{n-k}p^k}{1-(1-p)^n-p^n}.
\end{align}
Moreover, there are $n \choose k$ sequences with the same composition of zeros and ones and hence probability. Thus, the entropy is
\begin{align}\label{eqn: surj_eq_b}
    H(f(1),\ldots,f(n)) &= \sum_{k=1}^{n-1} {n \choose k} p_k \log{\tfrac{1}{p_k}}.
\end{align} 
\paragraph{Equality of functions up to a permutation of $\mathcal{N}$}\label{sec: equality_up_to_N_3}
Here we have equivalence classes of surjective functions up to a permutation of $\mathcal{N}$.

Here, after generating all the sequences and renormalizing the probabilities, we get $\stirling{n}{x}$ equiprobable sequences before considering groupings from equivalence relation on $\mathcal{N}$. Because of the surjectivity constraint, we have sequences of compositions, $(n_0,\ldots,n_{(x-1)})$ with $n_i \geq 1$. Hence the entropy can be computed as 
\begin{equation}\label{eqn: surj_n_u}
    H(f(1),\ldots,f(n)) = \sum_{(n_0,n_1,\ldots,n_{(x-1)}): \sum_{i}n_i = n} p(n_0,n_1,\ldots,n_{(x-1)})\log_2{\tfrac{1}{p(n_0,n_1,\ldots,n_{(x-1)})}},
\end{equation} 
where $p(n_0,n_1,\ldots,n_{(x-1)}) = \frac{1}{p_{norm}x^n} {n \choose {n_0..n_{(x-1)}}}$ with $p_{norm}$ being the normalizing constant, which we do not compute here. Note that this expression is the same as \eqref{eqn: any_n_u} except for the additional constraint here of $n_i \geq 1$ because of surjectivity. This expression is also computationally expensive like the one in \eqref{eqn: any_n_u}. Hence, we compute an upper bound to this entropy to get a better understanding of the behavior of this entropy as a function of $n$ and $x$.
For surjective functions with equality up to a permutation of $N$, we have ${n-1} \choose {n-x}$ equivalence classes, hence the entropy in this case can be upper bounded as
\begin{align}\label{eqn: surj_n_u_ub}
    H(f(1),\ldots,f(n)) \leq \log_2{{n-1} \choose {n-x}}.
\end{align}

For the Bern($p$) case, the composition of a sequence can be simply computed from the number of ones in them. There are $n \choose i$ sequences with the composition $(n_0,n_1)$, where $n_0\geq 1, n_1 \geq 1$ to satisfy surjectivity. Hence the entropy can be computed as
\begin{align}\label{eqn: surj_n_b}
    H(f(1),\ldots,f(n)) &= \sum_{k=1}^{n-1} p(k,n-k)\log_2{\tfrac{1}{p(k,n-k)}},
\end{align} 
where $p(k,n-k) = {n \choose k} \frac{(1-p)^{n-k}p^k}{1-(1-p)^n-p^n}$ with $(1-(1-p)^n-p^n)$ being the normalizing constant.

\paragraph{Equality of functions up to a permutation of $\mathcal{X}$}\label{sec: equality_up_to_X_3}
Now we consider surjective functions with equivalence up to permutation of $\mathcal{X}$. Consider the uniform distribution case. Observe that the total number of equivalence classes is equal to the number ways $\mathcal{N}$ into $x$ non-empty unlabelled subsets, which is given by $\stirling{n}{x}$. Moreover, observe that each class has exactly the same number of sequences. Hence, the grouped probabilities in this case is also uniform, which gives us the entropy as
\begin{align}\label{eqn: surj_x_u}
    H(f(1),\ldots,f(n)) &= \log_2{\stirling{n}{x}}.
\end{align} 

Now we take the Bern($p$) case. As $f$ is surjective, there are two invalid sequences: the all-zero sequence and the all-one sequence. Each equivalence class has exactly two sequences obtained by flipping the zeros and ones in them. Moreover, we have ${n \choose k}$ sequences with $k$ ones and $n-k$ zeros each occurring with probability $p^k(1-p)^{n-k}$. Consider $n$ to be odd, which gives
\begin{align}\label{eqn: surj_x_b}
    H(f(1),\ldots,f(n)) &= \sum_{k=1}^{\frac{n-1}{2}} {n \choose k}p(k,n-k)\log_2{\tfrac{1}{p(k,n-k)}},
\end{align} 
where $p(k,n-k) = \frac{p^k(1-p)^{n-k} + (1-p)^k p^{n-k}}{1-(1-p)^n-p^n}$. We can similarly compute the entropy for the case when $n$ is even, which we have omitted here for brevity.

\paragraph{Equality of functions up to permutations of $\mathcal{N}$ and $\mathcal{X}$}\label{sec: equality_up_to_N_and_X_3}

For surjective functions equivalence up to permutation of both $\mathcal{N}$ and $\mathcal{X}$, we first consider the uniform case. The computations are somewhat involved, hence we approximate the entropy with an upper bound
\begin{align}\label{eqn: surj_nx_u_ub}
    H(f(1),\ldots,f(n)) &\leq  \log_2{p_x(n)},
\end{align} 
since total number of equivalence classes is given by $p_x(n)$, where $p_x(n)$ represents the partitions of $n$ into $x$ parts.

Now we consider the Bern($p$) case. Take $n$ to be odd for simplicity. Each equivalence class exclusively consists of sequences with a fixed number of zeros or ones. Further, surjectivity implies that the all-zeros and the all-ones sequences are invalid. There are $\frac{n-1}{2}$ equivalence classes with the $k$th equivalence class consisting of sequences with exactly $k$ or $n-k$ ones in them. The probability of the $k$th class is given by $p(k,n-k) = {n \choose k}(p^k(1-p)^{n-k} + (1-p)^k p^{n-k})$ before the normalization. Hence the entropy is:
\begin{align}\label{eqn: surj_nx_b}
    H(f(1),\ldots,f(n)) &= \sum_{k=1}^{\frac{n-1}{2}} p(k,n-k)\log_2{\tfrac{1}{p(k,n-k)}},
\end{align} 
where $p(k,n-k) = {n \choose k}\frac{(p^k(1-p)^{n-k} + (1-p)^k p^{n-k})}{1-(1-p)^n-p^n}$.

\begin{figure}
\label{fig: comparison_3}
    \centering
    \subfloat[]{{\includegraphics[width=6cm]{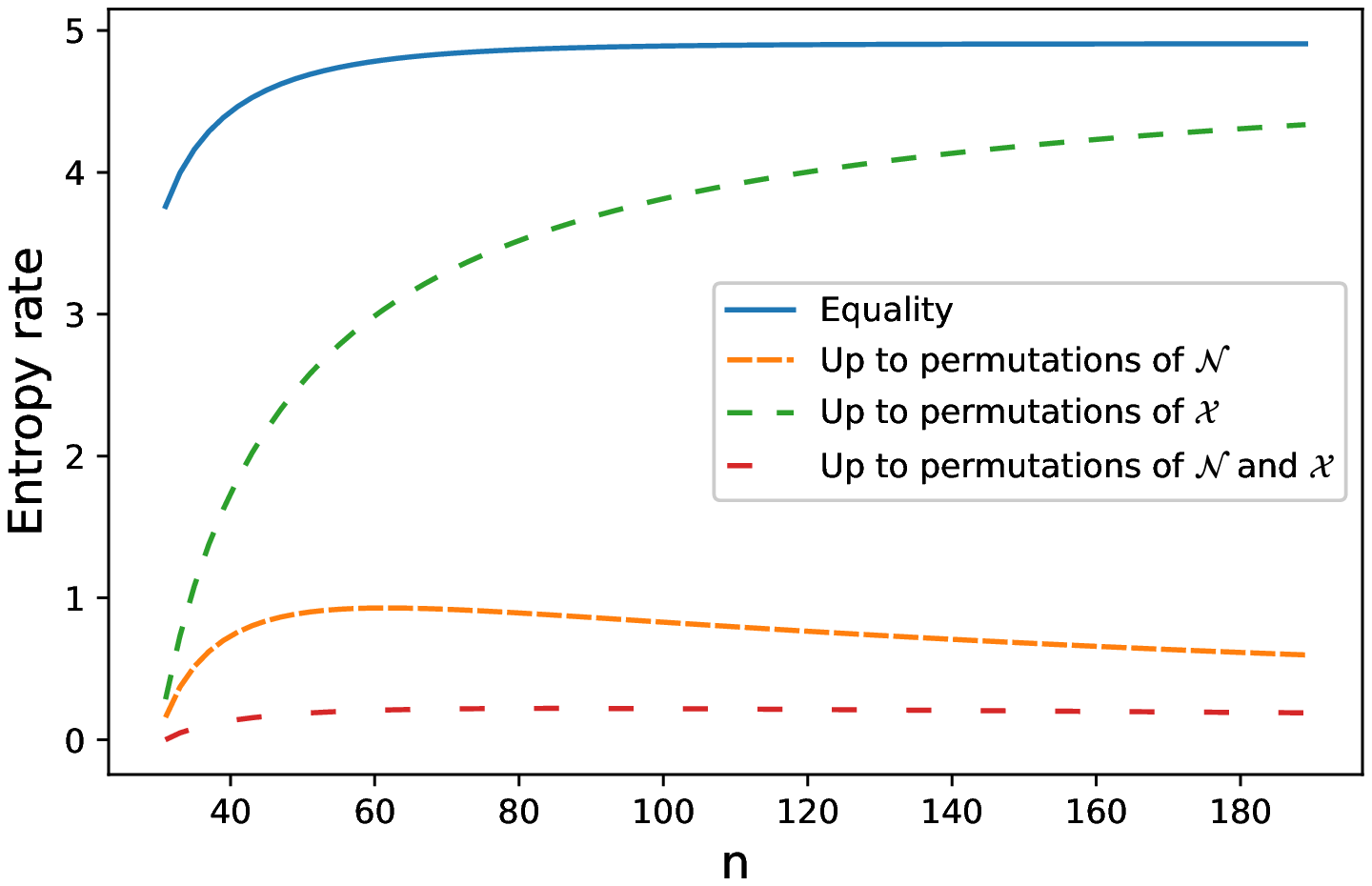} }\label{fig: N_U_C_3}}%
    \qquad
    \subfloat[]{{\includegraphics[width=6cm]{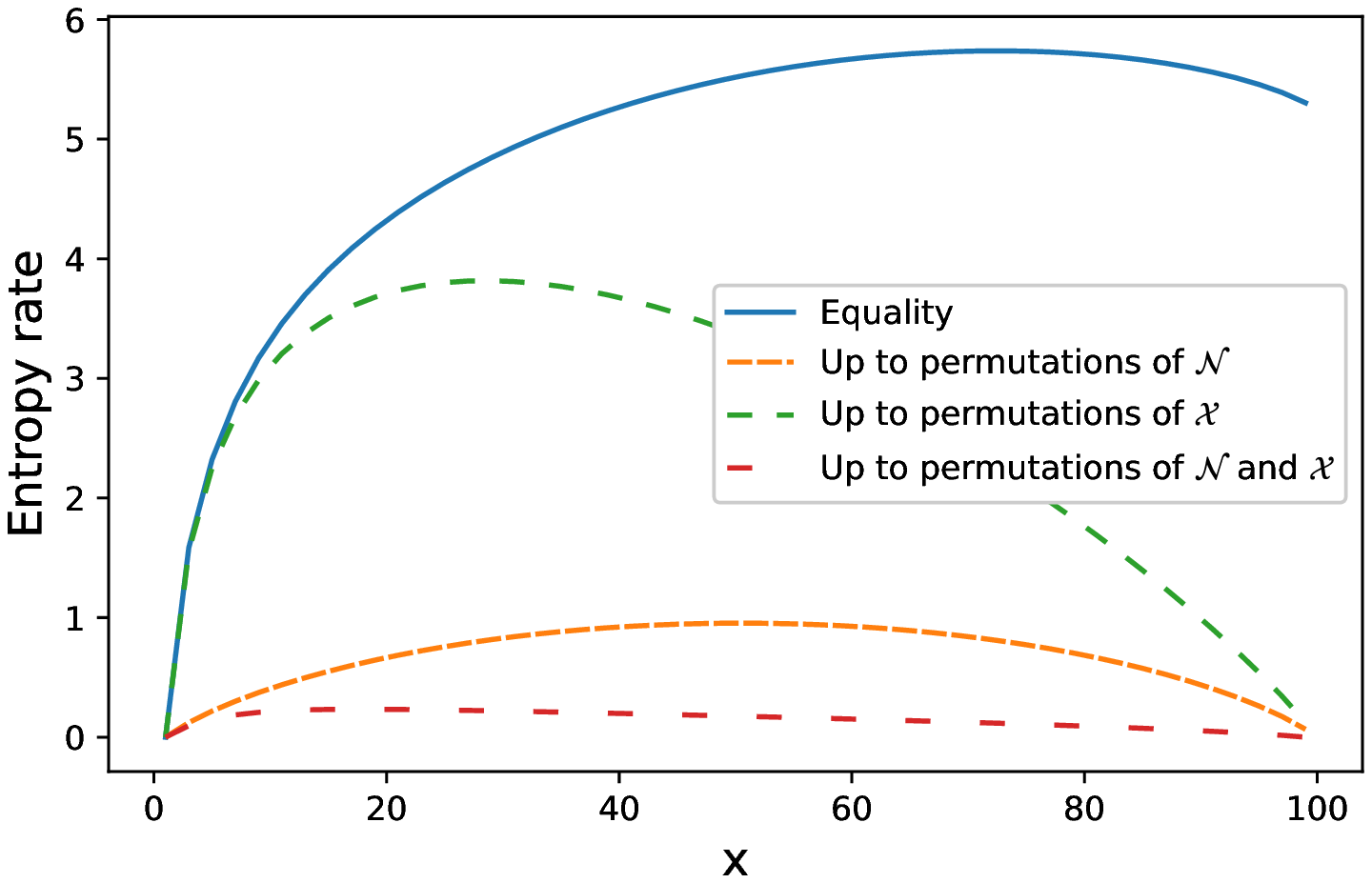} } \label{fig: X_U_C_3}}
    \qquad
    \subfloat[]{{\includegraphics[width=6cm]{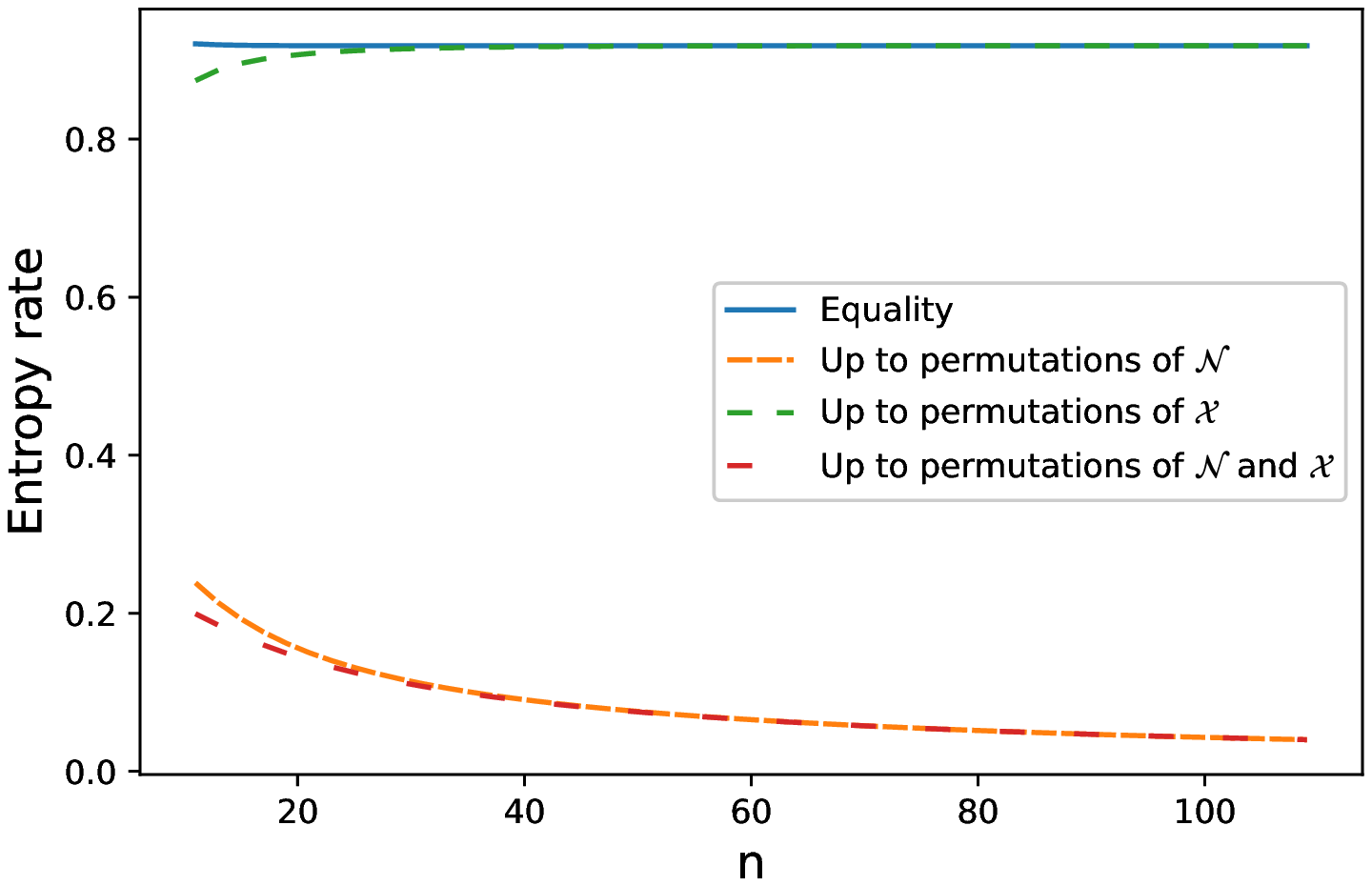} } \label{fig: N_B_C_3}}%
    \qquad
    \subfloat[]{{\includegraphics[width=6cm]{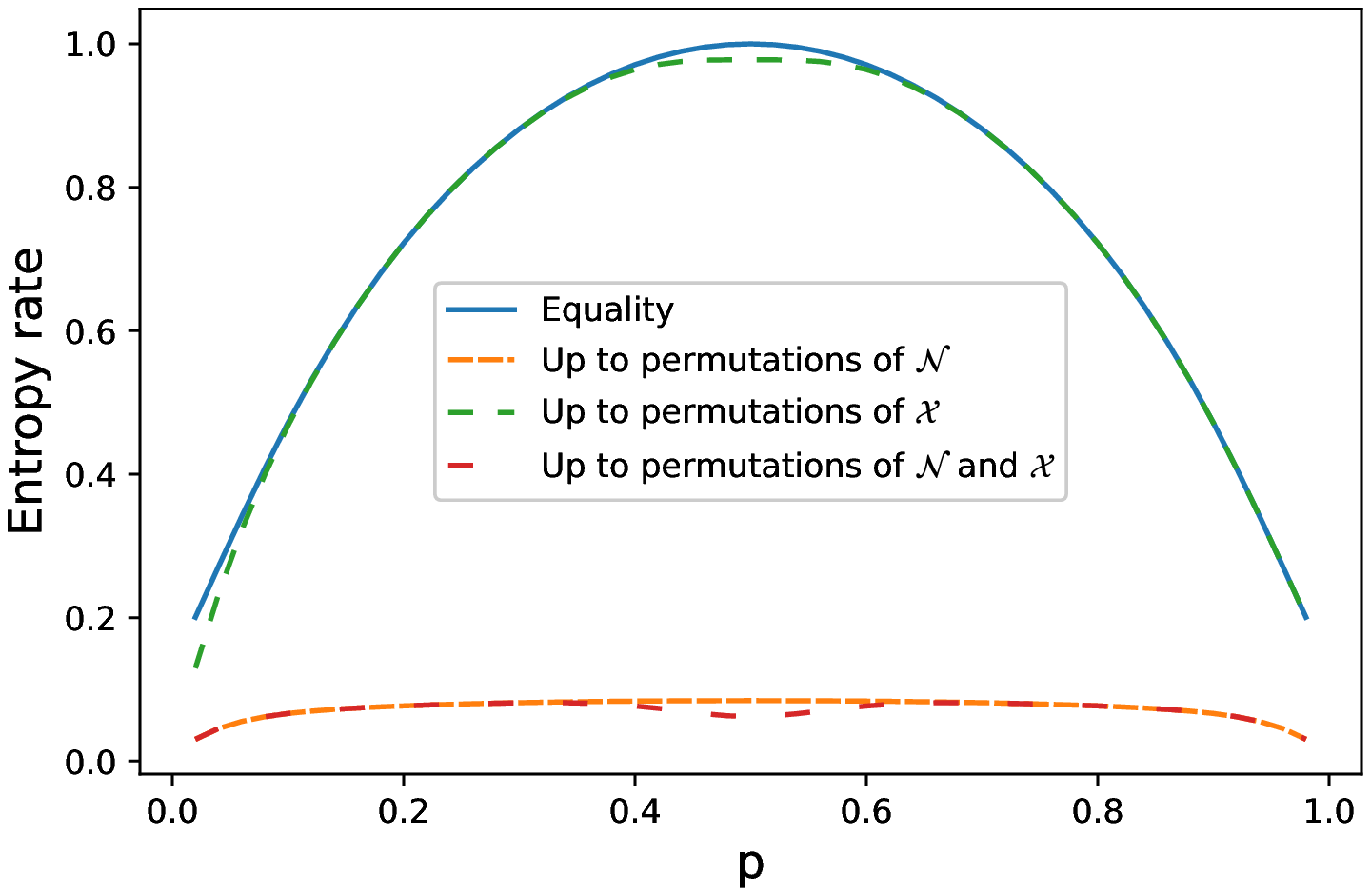} } \label{fig: P_B_C_3}}%
    \vspace{-3mm}
    \caption{Entropy rates for surjective functions vs. parameters $n,x,$ and $p$ for uniform and Bernoulli distributions and various equivalence relations.}%
\end{figure}

\paragraph{Comparison of entropies for surjective functions}\label{sec: comparison_3}
In Fig.~\ref{fig: N_U_C_3},~\ref{fig: X_U_C_3}, we plot entropy rates corresponding to \eqref{eqn: surj_eq_u}, \eqref{eqn: surj_n_u_ub}, \eqref{eqn: surj_x_u}, \eqref{eqn: surj_nx_u_ub} against $n$ with $x=30$ and $x$ with $n=100$ respectively. Note the significant difference in the entropy rates in the various cases. For surjective functions we have an implicit assumption $n\geq x$ which makes the entropy rates for equivalence in $\mathcal{N}$ much lower than for equivalence in $\mathcal{X}$. For the Bern($p$) case in Fig.~\ref{fig: N_B_C_3} and \ref{fig: P_B_C_3},  observe that main difference in entropy comes from equivalence in $\mathcal{N}$.
\section{Conclusion}
\label{sec:conclusion}
We have characterized the information-theoretic lossless compression limits for a wide range of non-sequentialities governed by the twelvefold way in combinatorics.  Going forward, it would be of interest to extend this approach to an even broader classification of group-theoretic invariances \cite{Steingart2012}, or at least enumerative combinatorics \cite[p.~57]{Bogart2004}.
It is also of interest to develop data structures that perform appropriate ``sorting'' to allow more efficient encoding and decoding of the twelvefold way than just enumerative source coding \cite{Cover1973}, cf.~\cite{Savari2004}.

\section*{Acknowledgment}
Thanks to Anand Sarwate for making us aware of the twelvefold way.

\Section{References}
\bibliographystyle{IEEEtran}
\bibliography{abrv,conf_abrv,refs}

\end{document}